\documentclass[iop,twocolappendix,apj]{emulateapj}
\usepackage{natbib}
\usepackage{graphicx}
\usepackage{hyperref}
\usepackage{breakurl}
\usepackage{apjfonts}
\usepackage{longtable}
\usepackage{gensymb}
\usepackage{float}
\newcommand{\msun} {$M_{\odot}$}

\newcommand{\Te} {$T_{\rm eff}$}
\newcommand{\logg} {$\log g$}
\newcommand{\loghe} {$\log$ (He/H)}

\newcommand{\psr} {PSR~J1738+0333}
\newcommand{\waspA} {WASP~J2047$-$25B}
\newcommand{\waspB} {WASP~J1628+10B}

\begin{document}

\slugcomment{Published in ApJ Letters}
\shortauthors{GIANNINAS ET AL}
\shorttitle{MIXED ATMOSPHERE PULSATING ELM WD PRECURSORS}

\title{DISCOVERY OF THREE PULSATING, MIXED-ATMOSPHERE, EXTREMELY LOW-MASS\\ WHITE DWARF PRECURSORS$^{*}$}


\author{A. Gianninas$^{1}$, Brandon Curd$^{1}$, G. Fontaine$^{2}$, 
Warren R. Brown$^{3}$, and Mukremin Kilic$^{1}$,}

\affil{$^{1}$Homer L. Dodge Department of Physics and Astronomy,
  University of Oklahoma, 440~W.~Brooks~St., Norman, OK 73019, USA;
  alexg@nhn.ou.edu}
\affil{$^{2}$D\'epartement de Physique, Universit\'e de Montr\'eal, C.P. 1628, 
  Succursale Centre-Ville, Montr\'eal, QC, H3C 3J7, Canada}  
\affil{$^{3}$Smithsonian Astrophysical Observatory, 60~Garden~St.,
  Cambridge, MA 02138, USA}

\begin{abstract}
We report the discovery of pulsations in three mixed-atmosphere,
extremely low-mass white dwarf (ELM WD, $M \leqslant$~0.3~\msun)
precursors. Following the recent discoveries of pulsations in both ELM
and pre-ELM WDs, we targeted pre-ELM WDs with mixed H/He atmospheres
with high-speed photometry. We find significant optical variability in
all three observed targets with periods in the range 320--590~s,
consistent in time-scale with theoretical predictions of $p$-mode
pulsations in mixed-atmosphere $\approx$~0.18~\msun\ He-core pre-ELM
WDs. This represents the first empirical evidence that pulsations in
pre-ELM WDs can only occur if a significant amount of He is present in
the atmosphere. Future, more extensive, timeseries photometry of the
brightest of the three new pulsators offers an excellent opportunity
to constrain the thickness of the surface H layer, which regulates the
cooling timescales for ELM WDs.  \\[0.1pt]
\end{abstract}


\footnotetext[*]{Based on observations obtained at the Gemini
  Observatory, which is operated by the Association of Universities
  for Research in Astronomy, Inc., under a cooperative agreement with
  the NSF on behalf of the Gemini partnership: the National Science
  Foundation (United States), the National Research Council (Canada),
  CONICYT (Chile), Ministerio de Ciencia, Tecnolog\'{i}a e
  Innovaci\'{o}n Productiva (Argentina), and Minist\'{e}rio da
  Ci\^{e}ncia, Tecnologia e Inova\c{c}\~{a}o (Brazil).}

\section{MOTIVATION}

Pulsating white dwarf (WD) stars provide a unique opportunity to probe
their internal structure by comparing the observed spectrum of
pulsation periods with the predictions of detailed asteroseismic models.
Several different families of pulsating WDs are known and they all
occupy fairly well defined instability regions in the \Te
--\logg\ plane \citep{fontaine08,winget08}. The largest class of
pulsating WDs are the so-called ZZ Ceti (or DAV) stars which are found
in a narrow range of effective temperature with 11,000~$\lesssim$
\Te\ $\lesssim$ 12,500~K \citep{gianninas11}. As these WDs cool and
enter the instability strip, they become unstable to non-radial
$g$-mode oscillations driven by a hydrogen partial ionization zone.

Recently, the discovery of pulsations in several extremely low-mass
(ELM) WDs has demonstrated that the canonical ZZ Ceti instability
strip extends to considerably lower surface gravities (5 $<$
\logg\ $<$ 7) \citep{hermes13c}. The ongoing ELM Survey
\citep[see][and references therein]{gianninas_ELM6,brown_ELM7}, has
led to the identification of seven pulsating ELM WDs
\citep{hermes12b,hermes13b,hermes13c,bell15} including the unique
system \psr\ \citep{kilic15}. These empirical results have been
complimented by the theoretical work of \citet{vg13},
\citet{corsico14,corsico16a}, and \citet{jeffery16}.

Recently, a pair of pulsating stars with evolutionary ties to ELM WDs
was discovered. The class of EL CVn-type binaries outlined in
\citet{maxted14a} are considered to be precursors to ELM WD
systems. The pre-ELM WDs in the EL CVn-type systems \waspA\ and
\waspB\ have been shown to pulsate \citep{maxted13,maxted14b}. These
pre-ELM WDs are located outside the boundaries of the extended ZZ Ceti
instability strip and thus their pulsations cannot be explained by the
usual driving mechanism ascribed to partially ionized
hydrogen. Instead, \citet{jeffery13} require significant amounts of
helium in the driving region to explain the observed $p$-mode
pulsations in these stars.

\citet{vg15} have shown that between the ZZ Ceti instability and the
V777 Her instability strip for helium atmosphere WDs, a continuum of
instability strips exist as the composition of the atmosphere is
varied from pure H to pure He. This phenomenon has recently been
explored in greater detail for the ELM WD regime by
\citet{corsico16b}.

The discovery of the pulsators in the two WASP systems and the
predictions of mixed-atmosphere pulsators outside the confines of the
traditional instability strips, spurred us to take a closer look at
objects from the ELM Survey where the presence of He in the atmosphere
had clearly been detected. In all, six candidates were
considered. These include J0751$-$0141, J1141+3850, J1157+0546,
J1238+1946, J1625+3632 \citep{gianninas14b}, and J0756+6704
\citep{gianninas_ELM6}. In Section~2 we describe the pulsation models
for these candidates and in Section~3 we present our timeseries
photometric observations. In Section~4, we discuss each of the newly
discovered pulsators individually and we conclude in Section~5 by
discussing the implications of these new discoveries and we consider
future avenues of research.

\begin{figure}[!h]
\centering
\includegraphics[scale=0.387,bb=28 164 582 653]{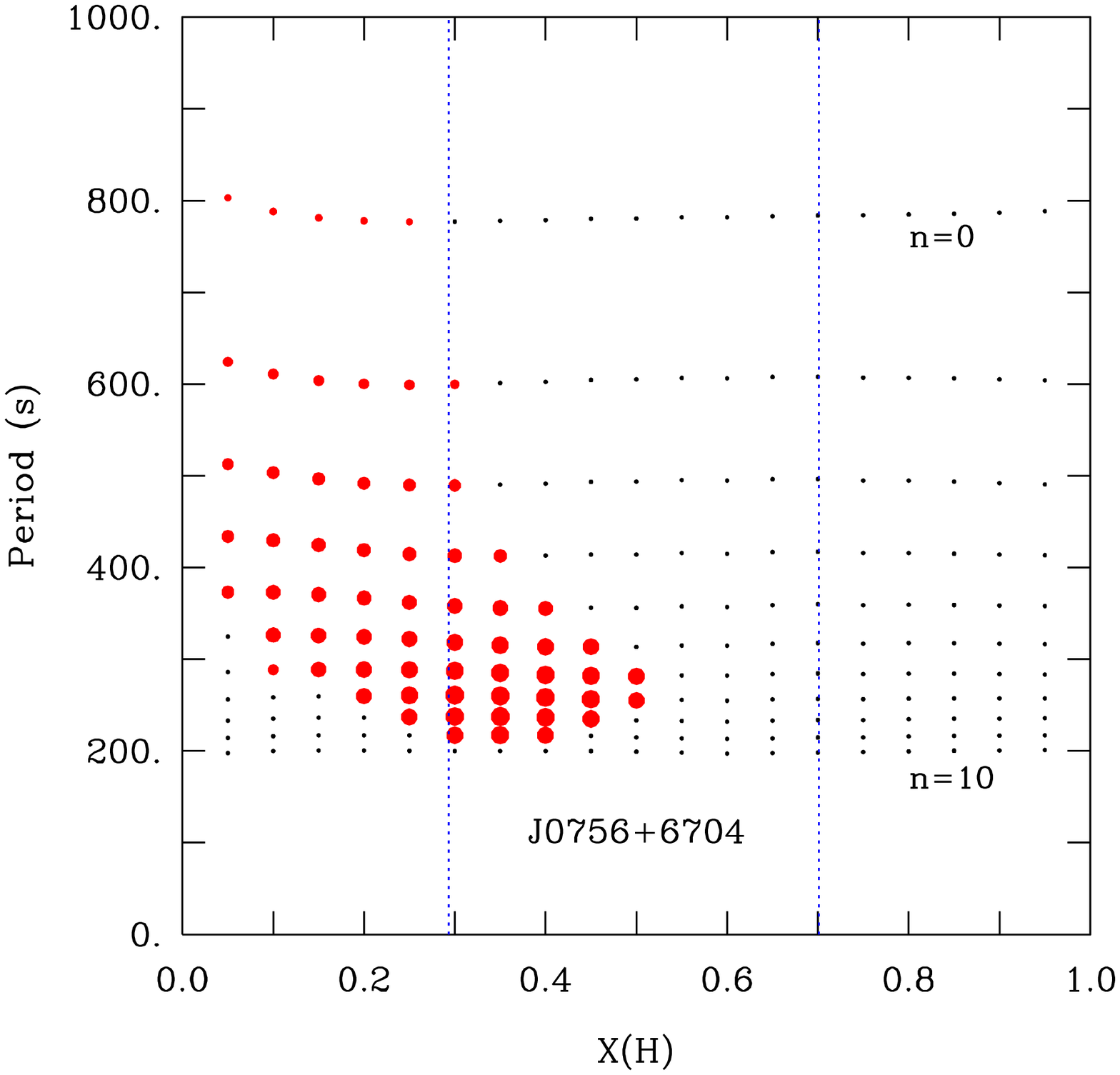}
\includegraphics[scale=0.387,bb=28 164 582 678]{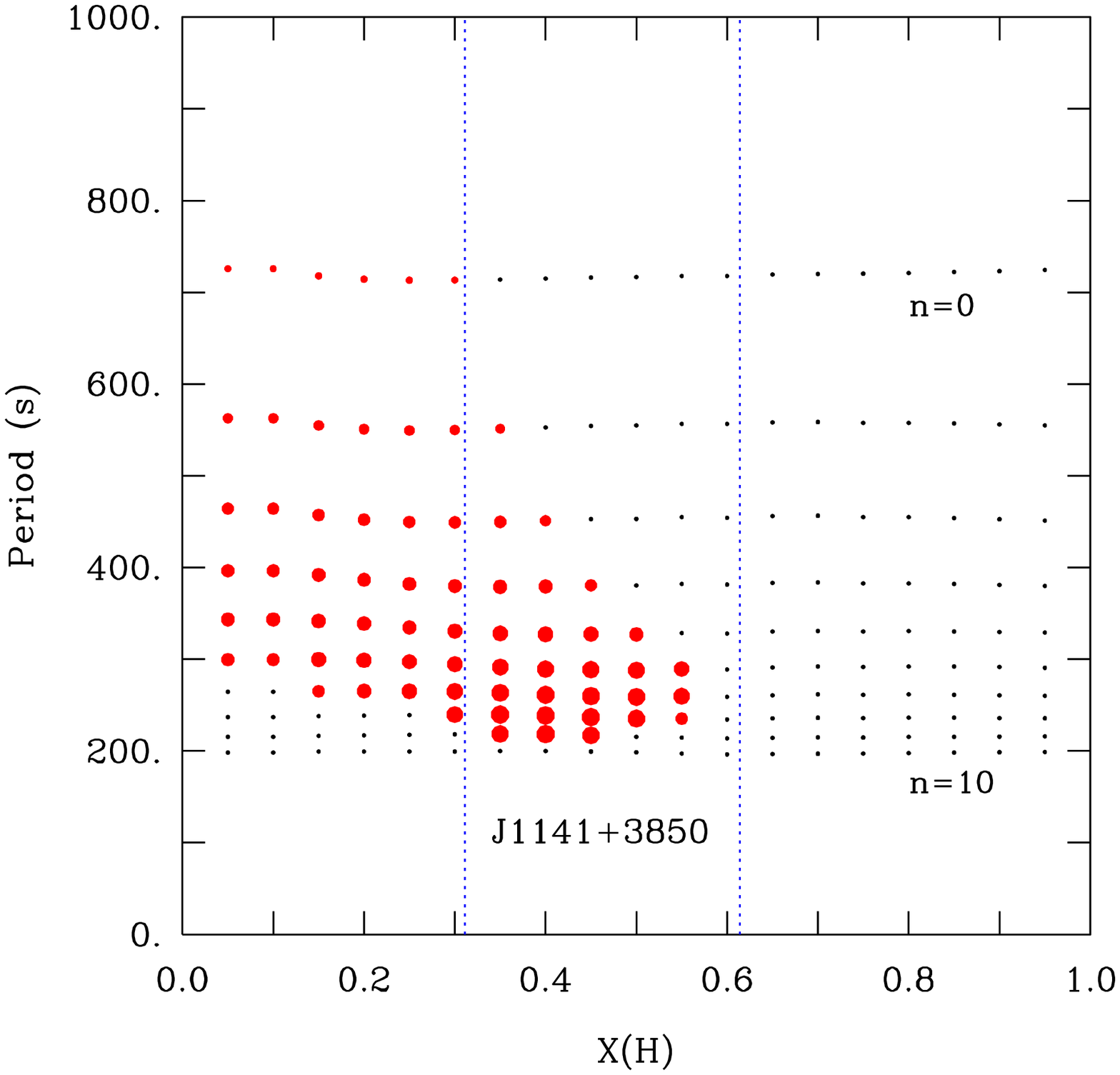}
\includegraphics[scale=0.387,bb=28 134 582 678]{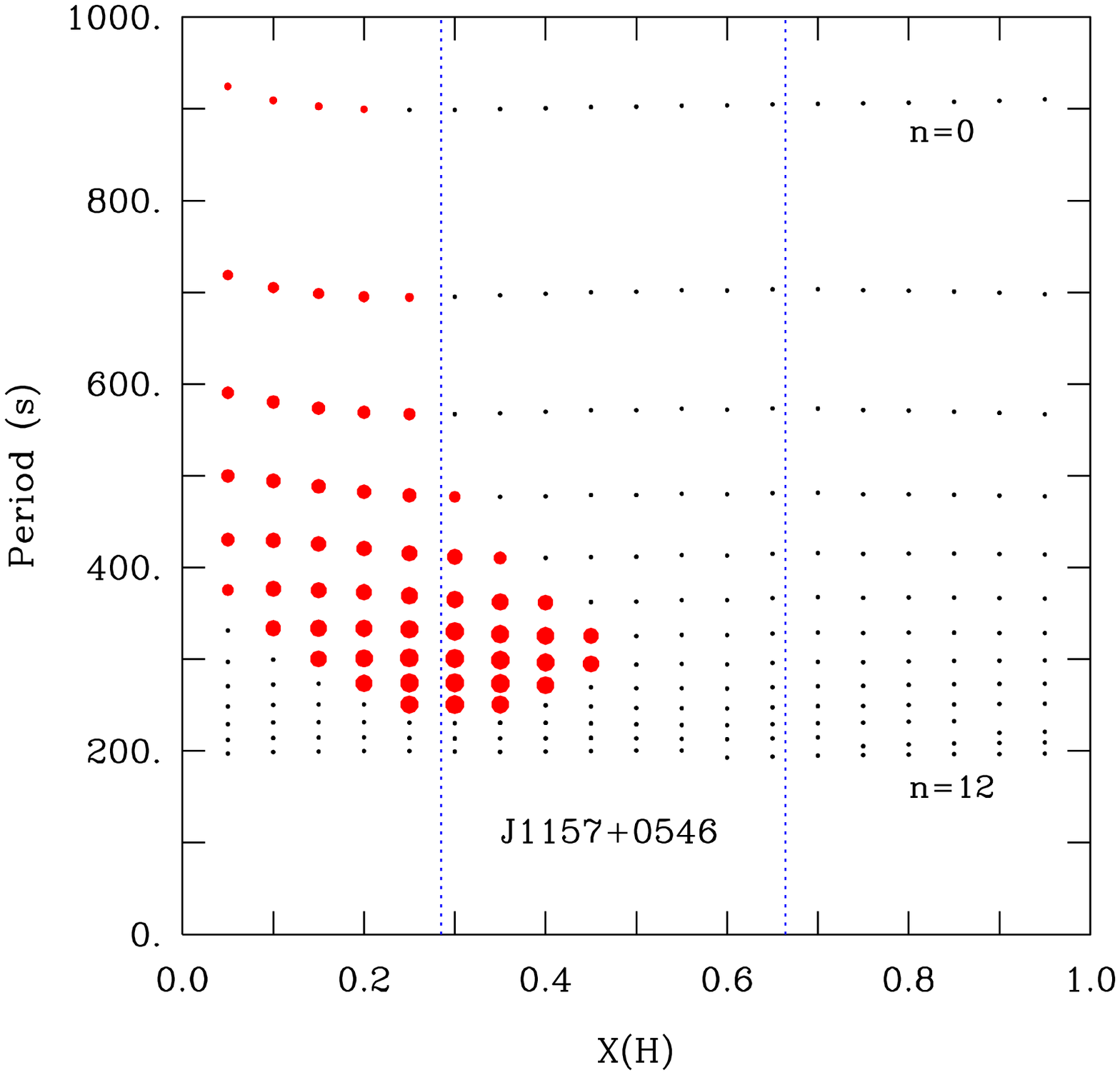}
\figcaption[f01a.eps]{Period spectrum for radial modes as a function
  of envelope composition for models of J0756 (top), J1141 (middle),
  and J1157 (bottom). Excited modes are shown as red dots while the
  black dots denote stable modes. The size of the red dots provides a
  logarithmic measure of the imaginary part of the eigenfrequency:
  larger dots correspond to more unstable modes. The blue lines
  indicate the range in composition based on the 1-$\sigma$
  uncertainties of the He abundance measured from the optical spectra.
  \label{fg:periods}}
\end{figure}

\section{Pulsation Models}

We carried out exploratory stability calculations with the help of the
Montr\'eal pulsation codes \citep{brassard92,fontaine94}. The
equilibrium models that we constructed are envelope models -- entirely
suitable for investigating the stability problem -- specified by the
fixed values of the effective temperature, surface gravity, mass, and
a variable value of the envelope composition (a varying mixture of H
and He). Mode driving, if any, is associated only with the partial
ionization of H and He as these models assume $Z = 0$. Some of our
results are depicted in Figure \ref{fg:periods}, where we show the
behavior of the spectrum of excited $p$-modes (with $\ell$ = 0 in this
specific example) as a function of the mass fraction of H in the
envelope. In all three cases, we find H/He mixtures that could
potentially lead to pulsations with 200 -- 800~s periods. None of the
$g$-modes investigated (extending in period up to 10,000~s) were found
to be excited in our models. In addition, no excitation -- $p$- or
$g$-mode -- was found in similar models for the three other
spectroscopic targets initially considered (i.e. J0751$-$0141,
J1238+1946, and J1625+3632).

The dotted blue lines correspond to the 1-$\sigma$ uncertainties on
the He abundance in the photosphere of each pre-ELM WD. Whether the He
is a product of a recent shell flash or has not yet settled out of the
atmosphere following the formation of the WD, it is reasonable to
assume that the He abundance increases as a function of depth in the
star and could thus be higher in the driving region. This would
effectively move the dotted blue lines to the left and make our models
more unstable to pulsations.

\section{OBSERVATIONS}

We obtained timeseries photometry of J0756+6704 ($g$~=~16.38~mag;
hereafter J0756) using the Smithsonian Astrophysical Observatory's
1.2~m telescope at the Fred Lawrence Whipple Observatory equipped with
KeplerCam \citep{sz05} on 2016 January 16. We obtained 282~$\times$~30~s
exposures over 4.2~hr. The chip was binned 2~$\times$~2 yielding a
$\approx$~22~s overhead per exposure and a plate scale of
0.67~arcsec~pixel$^{-1}$. Observations were obtained using an SDSS-$g$
filter under mixed conditions with a median seeing of
$\approx$~2.7~arcsec.

We obtained timeseries photometry of J1141+3850 ($g$~=~19.06~mag) and
J1157+0546 ($g$~=~19.82~mag) (hereafter J1141 and J1157, respectively)
using the 8~m Gemini-North telescope with the Gemini Multi-Object
Spectrograph \citep[GMOS, ][]{hook04} on 2016 February 17 as part of the
queue program GN-2016A-Q-59. To reduce the read-out time and the
telescope overhead to $\approx$~15~s, we binned the chip by
4~$\times$~4, resulting in a plate scale of 0.29 arcsec
pixel$^{-1}$. We obtained 289~$\times$~10~s exposures over 2~hr for
J1141 and 207~$\times$~20~s exposures over 2~hr for J1157. In both
cases, observations were obtained using an SDSS-$g$ filter. Conditions
were photometric with a median seeing of $\approx$~0.7~arcsec.

For reductions and calibrations we use the standard Image Reduction
and Analysis Facility (IRAF) routines and Gemini GMOS routines under
IRAF coupled with the daily bias and twilight sky frames. For J0756
(J1141 and J1157), we identify 22 (7, 4) non-variable reference stars
that are on the same CCD and amplifier as the target and use them to
calibrate the differential photometry.

\begin{figure*}[!ht]
\begin{center}
\centering
\includegraphics[scale=0.65,angle=-90,bb=22 -4 606 784]{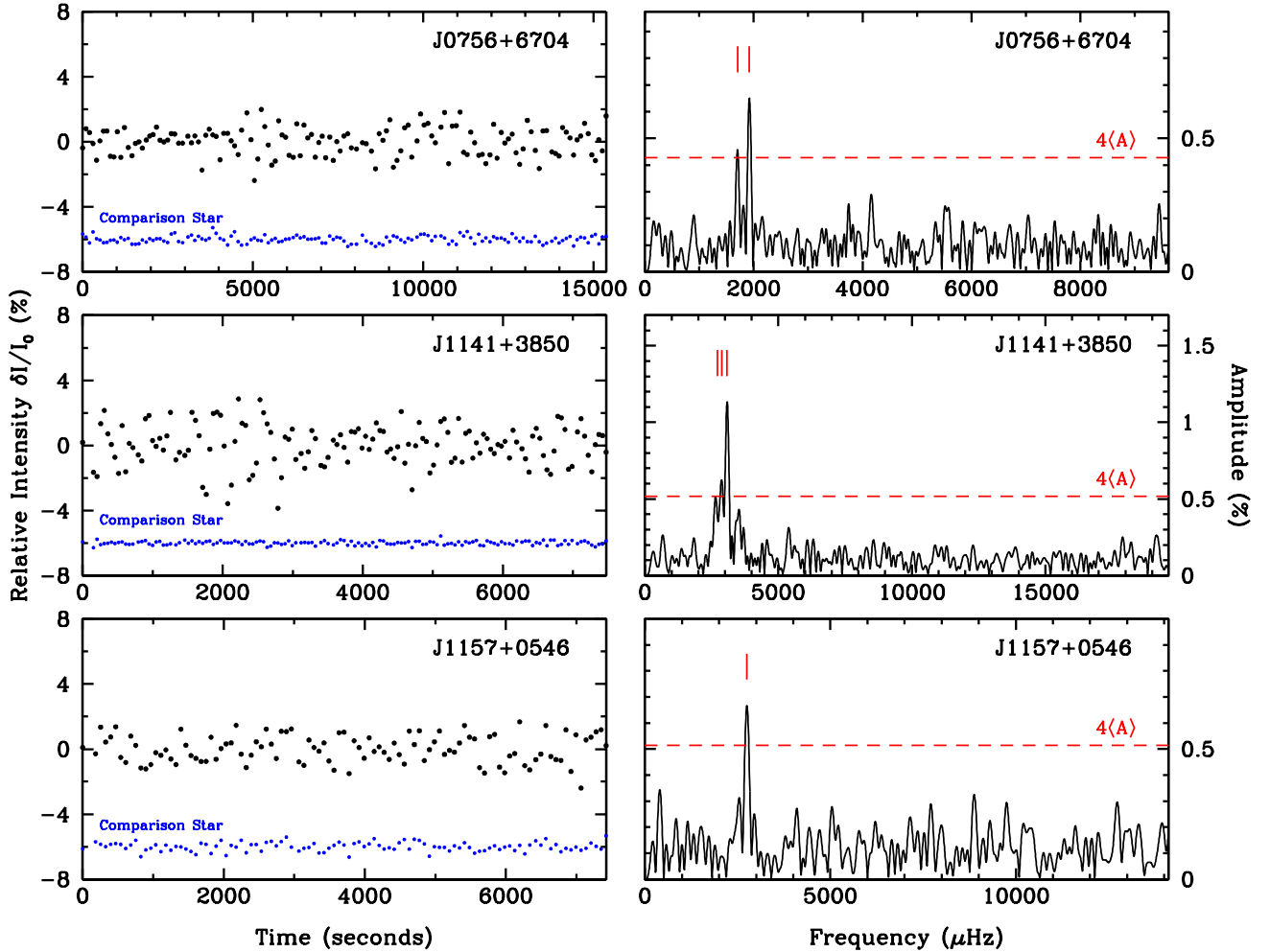}
\figcaption[f02.eps]{Left: observed light curves (black dots)
  for the three new pulsating pre-ELM WDs where a two-point boxcar
  smoothing has been applied. We also plot, in blue, the light curve
  of the brightest comparison star in the field, offset by $-$6~\% for
  clarity. Right: Fourier amplitude spectra of the observed
  light curves. The bandpass is plotted from 0 to the Nyquist
  frequency. We mark the 4$\langle A \rangle$ significance level as a
  dashed red line. The red tick marks denote the location of the
  significant frequencies.
\label{fg:lcft}}
\end{center}
\end{figure*}

\section{Photometric Analysis}

We used the \textsc{period04} package \citep{lenz05} to compute
Fourier transforms (FTs) from the light curves of all three targets
and extracted the significant pulsation frequencies.  Pulsation
frequencies are taken as significant if the amplitude exceeds
4$\langle A \rangle$ where $\langle A \rangle$ is the mean amplitude
of the FT from 0 to the Nyquist frequency.

In Figure~\ref{fg:lcft} we plot the discovery light curves for the
three new pulsating ELM WDs and also include the brightest comparison
star in the field for reference. We also display the FT for each
pulsator and identify the significant peaks in each case.

Table~\ref{tab:phot} summarizes the detected modes listing the period,
frequency, and relative amplitude. We also include the signal-to-noise
ratio (S/N) of each signal, calculated by comparing the mode amplitude
to the average FT amplitude after subtracting out the significant
periods of variability.

\subsection{J0756+6704}

\addtocounter{footnote}{2}

J0756 is a \Te\ =~11640~$\pm$~250~K, \logg\ =~4.90~$\pm$~0.14 pre-ELM
WD with \loghe\ $=-0.60~\pm~0.38$, $M$~=~0.181~$\pm$~0.011~\msun\ at a
distance of 1.6~$\pm$~0.3~kpc\footnote{Distance and mass estimates are
  based on the evolutionary models of \citet{althaus13};
  \citet{istrate14} models yield estimates that differ by less than
  5\% \citep{gianninas_ELM6}.}  \citep{gianninas_ELM6}.  It is found
in a 0.61871~days orbit with a companion that has a minimum mass of
0.82~$\pm$~0.03~\msun . The top panels of Figure~\ref{fg:lcft} show
the KeplerCam light curve and FT for J0756. We detect two significant
periods in J0756 at 521 and 587~s with relative amplitudes of 0.64\%
and 0.43\%, respectively. These periods are somewhat longer than
predicted by the models depicted in Figure~\ref{fg:periods}.

A preliminary exploration of a grid of appropriate pulsation models is
able to reproduce the two observed periods in J0756. The chosen model
has \Te\ =~11619~K, \logg\ =~4.86, an envelope composition of $X(\rm
H)$~=~0.395, an envelope mass of $\log M_{\rm H}/M_{\star}$~=~$-$1.87,
and assumes a convective efficiency of ML2/$\alpha$~=~1.0 \citep{vg13}, a
total mass of 0.186~\msun\ and a core of pure He.

The model reproduces the observed periods almost exactly with the
521~s period attributed to an $\ell$~=~0, $k$~=~2 mode and the 587~s
period corresponding to an $\ell$~=~2, $k$~=~0 mode. Both of these are
$p$-modes; the former is a radial mode while the latter is a
non-radial mode. However, this solution is not unique; we currently
have two detected periods to work with and these provide only modest
constraints on the models. Nonetheless, these periods are consistent
with those predicted by the model depicted in Figure~\ref{fg:periods},
albeit for the lowest value of $X$(H) within the 1-$\sigma$ range.

\begin{table}[!t]
\normalsize
\caption{Photometric Parameters for the New Pulsating ELM WDs}
\begin{center}
\begin{tabular*}{\hsize}{@{\extracolsep{\fill}}lr@{ $\pm$ }@{\extracolsep{0pt}}rr@{ $\pm$ }@{\extracolsep{0pt}}rr@{ $\pm$ }@{\extracolsep{0pt}}rc@{}}
\hline
\hline
\noalign{\smallskip}
SDSS & \multicolumn{2}{c}{Period} & \multicolumn{2}{c}{Frequency} & \multicolumn{2}{c}{Amplitude} & S/N \\
     & \multicolumn{2}{c}{(s)}    & \multicolumn{2}{c}{($\mu$Hz)} & \multicolumn{2}{c}{(\%)}      &     \\
\noalign{\smallskip}
\hline
\noalign{\smallskip}
J0756+6704 & 521 & 1 & 1921 & 4 & 0.64 & 0.08 & 6.6 \\
           & 587 & 2 & 1705 & 6 & 0.43 & 0.08 & 4.4 \\
\noalign{\smallskip}
\hline
\noalign{\smallskip}
J1141+3850 & 325 & 1 & 3073 &  7 & 1.12 & 0.10 & 9.8 \\
           & 346 & 2 & 2886 & 14 & 0.56 & 0.10 & 4.9 \\
           & 368 & 2 & 2719 & 16 & 0.51 & 0.09 & 4.5 \\
\noalign{\smallskip}
\hline
\noalign{\smallskip}           
J1157+0546 & 364 & 1 & 2751 & 11 & 0.66 & 0.10 & 5.5 \\
\noalign{\smallskip}
\hline
\end{tabular*}
\label{tab:phot}
\end{center}
\end{table}

\subsection{J1141+3850}

J1141 is a \Te\ =~11290~$\pm$~210~K, \logg\ =~4.94~ $\pm$~0.10 pre-ELM
WD with \loghe\ $= -0.53~\pm~0.27$, $M$~=~0.177~$\pm$~0.011~\msun\ at
a distance of 5.2~$\pm$~0.8~kpc \citep{gianninas14a}. It is found in a
0.25958~days orbit with a companion that has a minimum mass of
0.765~$\pm$~0.04~\msun. The middle panels of Figure~\ref{fg:lcft} show
the GMOS light curve and FT for J0756. In this case, we detect three
significant modes with periods of 325, 346, and 368~s and relative
amplitudes of 1.12, 0.56 and 0.51\%, respectively. Overall, the three
significant periods are shorter and have larger relative amplitudes
than the modes observed in J0756.

Of potential interest in the case of J1141 is the spacing of the
observed pulsation modes. The frequency spacings separating the three
modes are 187~$\pm$~16~$\mu$Hz and 167~$\pm$~21~$\mu$Hz, which agree
within the uncertainties. If these spacings are in fact equal, then it
is possible that we are observing a triplet produced by the rotational
splitting of an $\ell$~=~1 mode. If we assume this is correct, we can
estimate the rotation period of J1141. Taking a weighted mean of the
two spacings yields an average spacing of 178~$\pm$~13~$\mu$Hz which
corresponds to a relatively short rotational period of
1.6~$\pm$~0.1~hr.

We stress that rotational splitting is only one potential explanation
for the observed periods. It is more likely that we have detected
$p$-mode periodicities with different values of the degree
$\ell$. Indeed, the spacings between consecutive periods are too small
for all three modes to belong to the same degree value. On the other
hand, the period difference of 43~s between the two extreme values is
comparable to what theory suggests in that period range for two
adjacent modes with the same $\ell$ value, i.e., $\sim$40~s. This
would also account for the difference between the middle period of
346~s and the possible period of 384~s, both of them belonging to a
different value of $\ell$ than the pair 325-368~s.

\subsection{J1157+0546}

J1157 is a \Te\ =~11870~$\pm$~260~K, \logg\ =~4.81~$\pm$~0.13 pre-ELM
WD with \loghe\ $= -0.55~\pm~0.35$, $M$~=~0.186~$\pm$~0.011~\msun\ at
a distance of 9.2~$\pm$~1.9~kpc \citep{gianninas14a}. It is found in a
0.56500~days orbit with a companion that has a minimum mass of
0.46~$\pm$~0.05~\msun. The bottom panels of Figure~\ref{fg:lcft} show
the GMOS light curve and FT for J1157. We detect a single significant
periodicity at 364~s with a relative amplitude of 0.66~\%. This is
perfectly consistent with the predictions of the models shown in
Figure~\ref{fg:periods}. However, given that only a single period is
detected in J1157, it is impossible to draw any detailed conclusions
with regards to the exact pulsation mode that is observed.

\begin{figure}[!t]
\begin{center}
\centering
\includegraphics[scale=0.435,bb=30 97 592 654]{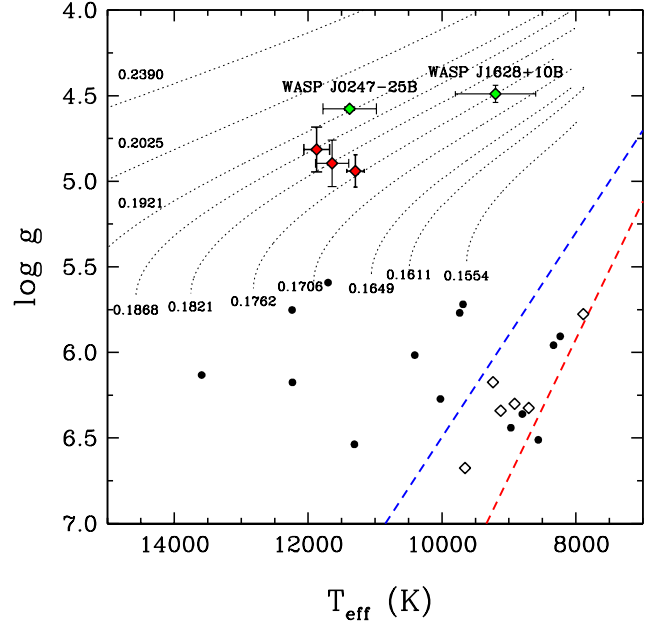}
\figcaption[f03.eps]{Region of the \Te--\logg\ plane containing the
  instability strip of pulsating ELM WDs (lower left), six of the
  known ELM pulsators (white diamonds), and ELM WDs where no pulsations
  are detected (black dots). The red diamonds represent the three new
  pulsators from this paper. The green diamonds correspond to the
  pulsators in EL CVn-type systems. The dashed lines denote the
  empirical blue and red boundaries of the instability strip
  \citep{gianninas_ELM6,tremblay15}. The dotted black lines represent the
  evolutionary tracks for He-core ELM WDs from \citet{althaus13} with
  stars evolving from the upper right to the lower left. The
  individual tracks are labeled by the final mass in \msun.
  \label{fg:strip}}
\end{center}
\end{figure}

\section{DISCUSSION}

\subsection{Three New Pulsating Pre-ELM WDs}

The discovery of these three new pulsating, mixed-atmosphere, pre-ELM
WDs represents yet another success of the predictive power of
non-adiabatic pulsation theory. In Figure~\ref{fg:strip} we plot the
location of the three new pulsators in the \Te--\logg\ plane along
with the previously known ELM WD pulsators which lie within the
confines of the extended ZZ Ceti instability strip. Given the
remarkable similarities in the atmospheric parameters of all three
targets, it could appear at first glance as if these pre-ELM WDs form
an entirely new class of pulsators.

Those stars are certainly different from the more compact pulsating
ELM WDs found in the ZZ Ceti instability strip illustrated in
Figure~\ref{fg:strip}. First, they are $p$-mode pulsators whereas the
latter are all $g$-mode variables with much longer periods\footnote{A
  possible detection of a $p$-mode was reported by \citet{hermes13b}
  in J1112+1117, but the evidence is marginal and has not been
  confirmed.}. Second, the driving process in the three objects
described here is dominated by the usual $\kappa$-mechanism---with
some contribution from convective driving---whereas the ZZ Ceti ELMs
are excited by convective driving at the base of a pure H ionization
zone. Third, the three stars of interest have mixed H/He
atmosphere-envelope compositions---the presence of He playing a
critical role in the driving process---whereas the ZZ Ceti ELMs are
excited by a mechanism that occurs in pure H layers.

On the other hand, the three new pulsators bear a strong similarity
with the other currently known pre-ELM WD pulsators found in the EL
CVn-type binaries \waspA\ and \waspB\ \citep{maxted13,maxted14a}. The
latter are also $p$-mode variables with periods in the same range as
those reported here and they occupy the same general region in the
spectroscopic Hertzprung--Russell diagram, i.e., on the contracting
branches of the evolutionary tracks (see Figure
\ref{fg:strip}). Indeed, there is a significant difference in the mean
densities of the new pre-ELM pulsators versus the pulsators in the EL
CVn systems. Based on our estimates of the mass and radius of the
pre-ELM WD pulsators from the \citet{althaus13} models, we calculate
mean densities of $\approx$~16, 19, 12 g/cm$^{3}$ for J0756, J1141,
and J1157, respectively. In contrast, the mean density for \waspA\ and
\waspB\ is $\approx$~5~g/cm$^{3}$. In addition, the companions in the
EL CVn systems are well-constrained through the observed eclipses and
that they are double-lined systems. It is clear that for both WASP
systems the companions are main sequence A stars. In contrast, the
pre-ELM WD binaries are all single-lined systems where the unseen
companion is assumed to be a cooler, more massive WD.

Because the spectra of the pulsating pre-ELM WD components in the EL CVn
systems are heavily polluted by the light of the A-star companions, it
has been difficult to infer their atmospheric He abundance; this
important quantity is still not available. If these objects belong to
the same class of pulsators as the three pre-ELM WDs reported in this
paper, He must necessarily be present in substantial quantity in the
atmospheres of the pre-ELM WD components of the EL CVn
systems. According to this scheme, the He content in \waspA\ should be
roughly comparable to the the values we need for explaining the
existence of the three pulsators discussed here, while it should be
significantly lower (by perhaps a factor of 2 by mass fraction) in the
cooler \waspB\ which would be part of a different instability strip
than the former. This situation is analogous to the continuum of
instability strips discussed by \citet{vg15} in the context of mixed
H/He atmosphere WDs of the GW Lib type.

\subsection{The Origin of Helium in Pre-ELM WDs}

Helium is the key ingredient that allows J0756, J1141, and J1157, as
well as the two WASP pulsators, to be unstable to
pulsations. Consequently, understanding why the He is present, and in
such significant quantity is fundamental if we are to explain why
these unique pre-ELM WDs exist.
 
The evolutionary sequences of \citet{althaus13} and \citet{istrate14}
predict that ELM WDs within a certain mass range will experience one
or more H shell flashes as they evolve. Indeed,
\citet{gianninas14a,gianninas14b} suggested that the observed He and
metals such as Ca were possibly the result of a recent shell
flash. The shell flash produces He and could also mix up the interior
of the star sufficiently to bring metals back to the surface. However,
the masses inferred for the three new pulsating pre-ELM WDs are below
the mass threshold for which shell flashes are expected. This is
particularly true for the \citet{istrate14} models, which require
$M$~>~0.21~\msun\ for shell flashes to occur.

If shell flashes are not the source of the He, then it is likely that
the He in these particular pre-ELM WDs is simply left over from the
formation of the WD itself and has not completely diffused out of the
atmosphere yet. If this scenario is correct, then our estimate of the
H layer mass is particularly useful. Our preliminary estimate of $\log
M_{\rm H}/M_{\star} = -1.87$ for J0756 implies a relatively thick
surface H layer, which is consistent with the evolutionary models of
\citet{althaus13} and \citet{corsico16b}.

\subsection{Future Prospects}

Extensive follow-up observations of these new mixed-atmosphere,
pulsating pre-ELM WDs will be useful in identifying all of their
periodicities. As J0756 is the brightest of the three targets, it
is well-suited for an extended observing campaign. Simultaneous
multicolor photometry in three or more bands would be particularly
useful for proper mode identification as the relationship between
pulsation amplitude and wavelength is directly related to the $\ell$
value of the observed mode. A detailed asteroseismic analysis of such
observations would help constrain the internal structure of these
pre-ELM WDs. As our preliminary analysis portends, the mass of H at
the surface could be constrained and lead to improved evolutionary
models.

\acknowledgements A.G. and M.K. gratefully acknowledge the support of
the NSF under grant AST-1312678, and NASA under grant NNX14AF65G. This
work was supported in part by the Natural Sciences and Engineering
Research Council of Canada. G.F. also acknowledges the contribution of
the Canada Research Chair Program. This work was supported in part by
the Smithsonian Institution.

{\it Facilities:} \facility{Gemini (GMOS), FLWO: 1.2m (KeplerCam)}


\end{document}